\documentclass{ifacconf}

\usepackage{graphicx}      
\usepackage{natbib}        

\usepackage{amsmath,amssymb}
\usepackage{tikz}
\usepackage{pgfplots}
\usepackage{placeins}
\pgfplotsset{compat=newest}
\newlength\breite
\newlength\hohe

\newcommand{\norm}[1]{\left\lVert#1\right\rVert}
\newcommand\numberthis{\addtocounter{equation}{1}\tag{\theequation}}
\newcommand{\refeq}[1]{\overset{#1}{=}}
\newcommand{\refleq}[1]{\overset{#1}{\leq}}

\newcommand{\refarrow}[1]{\overset{#1}{\Rightarrow}}

\newcommand{\MAS}{\mathcal O_\infty}
\newcommand{\contrMAS}{\mathcal O_\infty^\lambda}
\newcommand{\intMAS}{\mathcal{\bar O}_\infty^\lambda}
\newcommand{\setS}{\mathcal S_v}
\newcommand{\intS}{\mathcal{\bar S}_v}

\newcommand{\setE}{\mathcal{E}_\infty^\lambda}
\newcommand{\setY}{\mathcal Y}
\newcommand{\intY}{\mathcal{\bar Y}}

\renewcommand{\int}{\text{int }}

\makeatletter
\newcommand{\pushright}[1]{\ifmeasuring@#1\else\omit\hfill$\displaystyle#1$\fi\ignorespaces}
\newcommand{\pushleft}[1]{\ifmeasuring@#1\else\omit$\displaystyle#1$\hfill\fi\ignorespaces}
\makeatother

\begin{document}
\begin{frontmatter}
	
\vspace{-10ex}
	
{\textcopyright~2023 the authors. This work has been accepted to IFAC for publication under a Creative Commons Licence CC-BY-NC-ND}

\title{Online convex optimization for constrained control of linear systems using a reference governor\thanksref{footnoteinfo}} 

\thanks[footnoteinfo]{This work was supported by Deutsche Forschungsgemeinschaft (DFG, German Research Foundation) - 505182457.}

\author[First]{Marko Nonhoff} 
\author[Second]{Johannes K{\"o}hler} 
\author[First]{Matthias A. M{\"u}ller}

\address[First]{Leibniz University Hannover, Institute of Automatic Control, 30167 Hannover, Germany (e-mail: \{nonhoff,mueller\}@irt.uni-hannover.de).}
\address[Second]{ETH Z{\"u}rich, Institute for Dynamic Systems and Control, 8092 Z{\"u}rich, Switzerland (e-mail: jkoehle@ethz.ch)}

\begin{abstract}                
In this work, we propose a control scheme for linear systems subject to pointwise in time state and input constraints that aims to minimize time-varying and a priori unknown cost functions. The proposed controller is based on online convex optimization and a reference governor. In particular, we apply online gradient descent to track the time-varying and a priori unknown optimal steady state of the system. Moreover, we use a $\lambda$-contractive set to enforce constraint satisfaction and a sufficient convergence rate of the closed-loop system to the optimal steady state. We prove that the proposed scheme is recursively feasible, ensures that the state and input constraints are satisfied at all times, and achieves a dynamic regret that is linearly bounded by the variation of the cost functions. The algorithm's performance and constraint satisfaction is illustrated by means of a simulation example. 
\end{abstract}

\begin{keyword}
Optimal control, control of constrained systems, dynamic regret, online convex optimization, reference governor
\end{keyword}

\end{frontmatter}

\section{Introduction}
Application of online convex optimization (OCO) to the problem of controlling linear dynamical systems subject to time-varying cost functions has recently gained significant interest. In contrast to classical numerical optimization, in the OCO framework the cost functions are allowed to be time-varying and a priori unknown, compare, e.g., \cite{Shalev12,Hazan2016} and the references therein for an overview. More specifically, an OCO algorithm has to choose an action $u_t$ at every time instant $t$. Only after the action is chosen, the current cost function $L_t(u)$ is revealed to the algorithm, which leads to a cost $L_t(u_t)$ depending on the algorithm's chosen action. 

In the context of controller synthesis, the OCO framework is of interest due to its low computational complexity and ability to handle time-varying and a priori unknown cost functions. These kind of cost functions commonly have to be considered in practice, e.g., due to unknown renewable energy generation or a priori unknown energy prices, compare, e.g., \cite{Tang17}. Therefore, OCO-based controllers have been proposed for linear dynamical systems \citep{Li19,Nonhoff20} subject to disturbances \citep{Agarwal19} and in a purely data-driven setting with noisy output feedback \citep{Nonhoff22}. Such OCO-based control algorithms are typically analyzed theoretically in terms of dynamic regret, a performance measure adopted from the OCO framework. Dynamic regret is able to capture transient performance of the closed loop system and is defined as the difference between the accumulated closed-loop cost of the controller and some benchmark, which is typically defined in hindsight, i.e., with knowledge of all cost functions. Studying the dynamic regret of controllers for dynamical systems has recently gained increasing attention, compare, e.g., \cite{Dogan21, Didier22}. However, in the literature on OCO-based control, pointwise in time state and input constraints are only considered in \cite{Nonhoff21,Li21}, and restrictive assumptions or a limited setting are necessary in these works to guarantee constraint satisfaction at all times. In particular, \cite{Nonhoff21} invoke controllability arguments with a deadbeat controller which can deteriorate the practical performance, whereas \cite{Li21} consider constraint satisfaction under disturbances, but limit the setting to disturbance rejection, and, thus, only compare to controllers that drive the system to the origin. Therefore, we propose an OCO-based controller for constrained linear systems and time-varying, a priori unknown cost functions in this work, that does not rely on additional restrictive assumptions.

Another closely related line of research is so-called feedback optimization, where the goal is to control a dynamical system to the solution of a (possibly time-varying) optimization problem. Again, the main focus of research is on linear dynamical systems subject to disturbances \citep{Menta18, Lawrence18, Colombino20}, in the data-driven setting \citep{Bianchin21}, and for nonlinear systems \citep{Hauswirth21, Zheng20}. In contrast to the OCO-based approaches, typically only asymptotic guarantees in the form of stability of the closed-loop system are given. Moreover, in the feedback optimization setting constraints are only considered for the asymptotic steady state, while point-wise in time constraints on the state trajectory are generally not satisfied.

Motivated by the fact that constraints are ubiquitous in practice due to, e.g., actuator limitations or mechanical restrictions, and violation of these constraints can often be safety-critical, we aim to design an algorithm that can handle both time-varying and a priori unknown cost functions as well as pointwise in time state and input constraints. In order to address the shortcomings discussed above, we combine the OCO-framework with a reference governor (RG) that ensures satisfaction of constraints on both the inputs and states of the system at all times. Reference governors modify the reference command to a well-designed closed-loop system whenever application of the unmodified reference would lead to constraint violation, compare, e.g., \cite{Garone17} for a recent survey on the topic. More specifically, we use online projected gradient descent \citep{Zinkevich03}, a well-studied method from OCO, to track the time-varying optimal steady state of the system under control. Then, comparable to the approach taken in \cite{Kalabic14}, we design a reference governor based on a $\lambda$-contractive set to enforce constraint satisfaction. Additionally, our approach using a $\lambda$-contractive set ensures a minimal rate of progress at each time step, which finally guarantees a linearly bounded dynamic regret. To the authors' best knowledge, this is the first work that proves bounded dynamic regret for a reference governor scheme.

This paper is organized as follows. Section~\ref{sec:setting} presents the setting considered in this paper. In Section~\ref{sec:algorithm}, we introduce and explain the proposed control scheme, and in Section~\ref{sec:results} we prove theoretical guarantees for the emerging closed loop, in particular recursive feasibility, constraint satisfaction at all times, and a bound on the dynamic regret. A numerical simulation example is given in Section~\ref{sec:example}. Section~\ref{sec:conclusion} concludes the paper.

\textit{Notation:} We denote the set of integer numbers greater than or equal to zero by $\mathbb N_{\geq0}$. For a vector $x\in\mathbb R^n$, $\norm{x}$ is the Euclidean norm. For a matrix $A\in\mathbb R^{n\times m}$, the corresponding induced matrix 2-norm is $\norm{A}$ and $\rho(A)$ denotes its spectral radius. The identity matrix of size $n\times n$ is given by $I_n$ and $0_{m\times n} \in \mathbb R^{m \times n}$ is the matrix of all zeros. The gradient of a function $f:\mathbb R^n \rightarrow \mathbb R$ evaluated at $x\in\mathbb R^n$ is denoted by $\nabla f(x)$. For two sets $\mathcal S, \mathcal T \subseteq \mathbb R^n$, $\mathcal S \oplus \mathcal T := \{x+y: x\in\mathcal S,y\in\mathcal T\}$ is the Minkowski set sum, the interior of $\mathcal S$ is $\int\mathcal S$, and for a compact set $\mathcal S\subseteq\mathbb R^n$, $\Pi_{\mathcal S}(x):=\arg\min_{s\in\mathcal S}\norm{x-s}$ denotes projection of the point $x\in\mathbb R^n$ onto the set $\mathcal S$.

\section{Setting} \label{sec:setting}

We consider linear time-invariant systems of the form
\begin{align}
	x_{t+1} = A x_t + Bu_t, \label{eq:sys}
\end{align}
with some initial state $x_0\in\mathbb R^n$, where $x\in\mathbb R^n$ is the system state, $u\in\mathbb R^m$ the system input, and $t\in\mathbb N_{\geq0}$. System~\eqref{eq:sys} is subject to constraints
\begin{align}
	y_t = C_0x_t + D_0u_t \in \setY \subseteq \mathbb R^m, \label{eq:constraints}
\end{align}
which have to be satisfied at all times $t\in\mathbb N_{\geq0}$.

\begin{assum} \label{assump:sys}
	The pair $(A,B)$ is stabilizable.
\end{assum}

\begin{assum} \label{assump:Y}
	$\mathcal Y$ is compact, convex, and $0 \in \text{int}~\mathcal Y$.
\end{assum}

The goal is to optimize performance measured in terms of time-varying, a priori unknown, convex cost functions $L_t(u_t,x_t)$. More specifically, at each time $t$, only the previous cost functions $L_0, \dots, L_{t-1}$ are known and we want to solve the optimization problem
\begin{equation*}
	\min_{u} \sum_{t=0}^T L_t(u_t,x_t) \quad
	\text{s.t. } \text{\eqref{eq:sys} - \eqref{eq:constraints}}.
\end{equation*}
By Assumption~\ref{assump:sys}, we can design a linear state feedback $K\in\mathbb R^{m\times n}$ such that $A_K:=A+BK$ is Schur stable, i.e., $\rho(A_K)$ is strictly smaller than one. Then, we can define $u_t = v_t + Kx_t$ and rewrite the system dynamics~\eqref{eq:sys} and constraints~\eqref{eq:constraints} as
\begin{subequations} \label{eq:sys2}
	\begin{align}
		x_{t+1} &= A_K x_t + Bv_t \label{eq:sys2_dynamics} \\
		y_t &= C x_t + D v_t \in \setY, \label{eq:sys2_constraints}
	\end{align}
\end{subequations}
where $C=(C_0 + D_0K)$ and $D = D_0$.
Similarly, the time-varying optimization problem can be equivalently reformulated as
\begin{equation}	
		\min_{v} \sum_{t=0}^T L_t(v_t+Kx_t,x_t) \quad
		\text{s.t. } \text{\eqref{eq:sys2}}. \label{eq:opt_problem}
\end{equation}
Let $S_K = (I_n - A_K)^{-1}B$ be the map\footnote{Since $A_K$ is Schur stable, the inverse exists and the map is unique.} from an input to the corresponding steady state of the stabilized system~\eqref{eq:sys2} and define the set of all feasible steady-state inputs as $\setS := \{v: (CS_K+D)v\in\mathcal Y\}$.

\begin{assum} \label{assump:L}
	The cost functions $L_t(u_t,x_t)$ are Lipschitz continuous with Lipschitz constant $l_L$ for all $t\in\mathbb N_{\geq0}$, i.e., $L_t(u_t,x_t) - L_t(\tilde u_t, \tilde x_t) \leq l_L \norm{(u_t,x_t) - (\tilde u_t, \tilde x_t)}$ holds for all $(u_t,x_t),~(\tilde u_t,\tilde x_t) \in \mathcal Z := \{(u,x)\in\mathbb R^{m}\times\mathbb R^n: C_0x+D_0u\in\mathcal Y\}$, and the steady-state cost functions $L^s_t(v) = L_t(v+KS_Kv,S_Kv)$ are $\alpha_v$-strongly convex and $l_v$-smooth\footnote{Compare \cite[Definition 2.1.3 and (2.1.9)]{Nesterov18}.} for all $t\in\mathbb N_{\geq0}$ and $v\in\setS$.
\end{assum}

Assumption~\ref{assump:L} is a common assumption in the literature on OCO-based control, compare, e.g., \cite{Li19,Nonhoff22}.

Since the cost functions $L_t$ are a priori unknown, we can in general not compute the minimizing input of~\eqref{eq:opt_problem} online. Instead, similar to \cite{Nonhoff22}, we adopt the strategy of tracking the a priori unknown and time-varying optimal steady-state reference given by
\begin{equation}
	\eta_t := \arg\min_{r}~L^s_{t}(r) \quad \text{s.t. } r \in \intS. \label{eq:def_eta}
\end{equation}
where $\intS$ is a compact, convex subset of $\setS$ such that $\intS \subseteq \int \setS$. The optimal steady state can be recovered by $\theta_t := S_K\eta_t$. Then, we define dynamic regret $\mathcal R$ as the difference between the accumulated closed-loop cost and the optimal steady-state cost as
\begin{equation}
	\mathcal R := \sum_{t=0}^T L_t(v_t + Kx_t,x_t) - L^s_t(\eta_t). \label{eq:def_regret}
\end{equation}
The goal is to achieve a bound for the dynamic regret~$\mathcal R$ that is linear in the path length given by $\sum_{t=1}^T\norm{\eta_t - \eta_{t-1}}$, because \cite{Li19} show for a similar, unconstrained setting that the best achievable bound is linear in the path length. Additionally, \cite{Nonhoff22a} prove that such a linear bound implies asymptotic stability under mild assumptions in the unconstrained case.

\section{OCO-based control using a reference governor} \label{sec:algorithm}
\subsection{Design of the reference governor}

In order to ensure satisfaction of the pointwise in time state and input constraints in~\eqref{eq:constraints}, we design a reference governor in this section. Reference governors compute a feasible reference command $v_t$ at each time~$t$ such that, if $v_t$ is applied constantly to the system~\eqref{eq:sys2}, then the constraints $y_t \in \setY$ are satisfied for all future time steps. This can be achieved using the maximal output admissible set (MAS) \citep{Gilbert91}.

\begin{defn} \label{def:MAS}
	The maximal output admissible set of a system $x_{t+1} = Ax_t$ with constraints $Cx_t \in \mathcal Y$ is defined as $\MAS = \{ x \in \mathbb R^n:~CA^tx \in \setY~\forall t \in \mathbb N_{\geq0} \}$.
\end{defn}

Note that any MAS $\MAS$ is positively invariant by definition, i.e., $A\MAS \subseteq \MAS$. Moreover, the MAS (or a close inner approximation thereof) can be calculated efficiently if $\mathcal Y$ is polytopic, $A$ is at least Lyapunov stable, and the pair $(A,C)$ is observable \citep{Gilbert91}.

Similar to the approach presented in \cite{Kalabic14}, we employ a $\lambda$-contractive set in our reference governor in order to ensure a sufficient rate of convergence and prove bounded dynamic regret.

\begin{defn} \label{def:contr_set}
	For a system $x_{t+1} = Ax_t$, a set $\mathcal X \subseteq \mathbb R^n$ is $\lambda$-contractive with some $\lambda\in(0,1)$ if it is compact, convex, $0\in\int\mathcal X$, and $A\mathcal X \subseteq \lambda\mathcal X$.
\end{defn}

We denote by $e_t := x_t - S_Kv_t$ the error between the state $x_t$ and the steady state corresponding to the reference $v_t$. Consider a constant reference $v_t = v$ for all $t\in\mathbb N_{\geq0}$. Then, the error dynamics of the system~\eqref{eq:sys2_dynamics} and the constraints~\eqref{eq:sys2_constraints} written in the error coordinates are
\begin{subequations} \label{eq:error_dynamics_constant_v}
\begin{align}
	&e_{t+1} = x_{t+1} - S_Kv = A_K x_t - A_KS_Kv = A_Ke_t, \label{eq:error_dynamics_constant_v_dynamics} \\
	&y_t = C(e_t + S_Kv) + Dv = Ce_t + (CS_K+D)v \in \mathcal Y.
\end{align}
\end{subequations}
Hence, we choose a $\lambda$ satisfying $\rho(A_K) < \lambda < 1$ and define $\contrMAS$ as the MAS of
\begin{subequations} \label{eq:sys_MAS}
	\begin{align}
		\nu_{t+1} &= \nu_t, \\
		\chi_{t+1} &= \frac{1}{\lambda} A_K\chi_t, \\
		\psi_t &= C\chi_t + (CS_K+D)\nu_t \in \setY. \label{eq:sys_MAS_constr}
	\end{align}
\end{subequations}
Since $\setY$ is compact, convex, and $0\in\int\setY$ by Assumption~\ref{assump:Y}, $\contrMAS$ is closed, convex, and $0\in\text{int }\contrMAS$ \citep{Gilbert91}. Moreover, $\contrMAS$ has the following properties.

\begin{lem} \label{lem:properties_contrMAS}
	Let Assumptions~\ref{assump:sys} and~\ref{assump:Y} be satisfied and let $\setE(v) := \{ e\in\mathbb R^n: (v,e) \in \contrMAS \}$. For any constant input $v_t = v \in \setS$ for all $t\in\mathbb N_{\geq0}$, (i) $\setE(v)$ is $\lambda$-contractive for~\eqref{eq:error_dynamics_constant_v_dynamics}, and (ii) $y_t \in \mathcal Y$ for all $t\in\mathbb N_{\geq0}$ if $x_0 \in \setE(v) \oplus \{S_Kv\}$.
\end{lem}
\begin{pf}
	(i) Since the MAS is positively invariant, we have by definition of $\contrMAS$ that
	\begin{align*}
		&e_t \in \setE(v) \Leftrightarrow \left(v,e_t\right) \in \contrMAS \Rightarrow \left(v,\frac{1}{\lambda} A_K e_t \right) \in \contrMAS \\
		\Leftrightarrow~&e_{t+1} = A_K e_t \in \lambda \setE(v). \numberthis  \label{eq:lambda_contr_error}
	\end{align*}
	(ii) For any $v\in\setS$, we have $v \in \setS \Leftrightarrow (CS_K+D)v \in \setY \Leftrightarrow (v,0) \in \contrMAS \Leftrightarrow 0 \in \setE(v)$, i.e., $0\in\setE(v)$ for any $v\in\setS$. By convexity of $\contrMAS$, this implies $\lambda\setE(v) \subseteq \setE(v)$. By~\eqref{eq:lambda_contr_error}, we get for any $v\in\setS$ and $e_t\in\setE(v)$, $e_t\in\setE(v) \Rightarrow e_{t+1} \in \lambda\setE(v)\subseteq \setE(v)$, i.e., the set $\setE(v)$ is positive invariant for the system~\eqref{eq:error_dynamics_constant_v_dynamics}. Since $x_0 \in \setE(v) \oplus \{S_Kv\}$ implies $e_0 = x_0-S_Kv \in \setE(v)$, we have $e_t \in \setE(v)$ for all $t\in\mathbb N_{\geq0}$. Moreover, $e_t \in \setE(v)$ and the fact that $\contrMAS$ is the MAS of system~\eqref{eq:sys_MAS} with output~\eqref{eq:sys_MAS_constr}, imply $e_t \in \,\setE(v) \Rightarrow y_t = Ce_t + (CS_K+D)v \in \setY$. \hfill $\square$
\end{pf}

Lemma~\ref{lem:properties_contrMAS} shows that, if the reference input $v_t$ is kept constant and the error $e_t = x_t - S_Kv_t$ together with the reference $v_t$ is contained in $\contrMAS$, then we can ensure contraction of the error and satisfaction of the constraints~\eqref{eq:constraints}.

\begin{figure} 
	\begin{center}
		\small
		\begin{tikzpicture}[node distance = 50,scale = 1]
	\clip (0,.5) rectangle (9,-1.6);
	\tikzstyle{block} = [draw, minimum size=.75cm, thick, align = center]
	\tikzstyle{add} = [draw, shape = circle, inner sep = 0pt, minimum size=.35cm, thick]
	\tikzstyle{circle} = [draw, shape=circle, inner sep = 0pt, minimum size = .15cm, fill=black, ultra thick]
	
	
	\coordinate (start) at (0,0);
	\node[block] (OCO) [right of = start, xshift = -10] {OCO};
	\node[block] (RG) [right of = OCO] {RG};
	\node[add] (add) [right of = RG, xshift = -15] {};
	\node[block] (system) [right of = add,xshift = 5] {$x_{t+1} = Ax_t + Bu_t$};
	\node[circle] (circ) [right of = system] {};
	\node[block] (K) [below of = system, yshift = 25] {$K$};
	\node[circle] (circ2) [below of = circ, yshift = 25] {};
	\coordinate[right of = circ, xshift = -40] (end);
	\coordinate[below of = circ2, yshift = 35] (aux);
	
	\draw[->] (start) -- (OCO) node[above,midway] {$L_{t-1}$};
	\draw[->] (OCO) -- (RG) node[above,midway] {$r_t$};
	\draw[->] (RG) -- (add) node[above,midway] {$v_t$};
	\draw[->] (add) -- (system) node[above,midway] {$u_t$};
	\draw[->] (system) -- (end) node[above,midway] {$x_t$};
	\draw (circ) -- (circ2) -- (aux);
	\draw[->] (circ2) -- (K);
	\draw[->] (K) -| (add);		
	\draw[->] (aux) -| (RG);
\end{tikzpicture}
	\end{center}
	\vspace{-3ex}
	\caption{Block diagram of the OCO-RG scheme}
	\label{fig:block_diagram}
\end{figure}
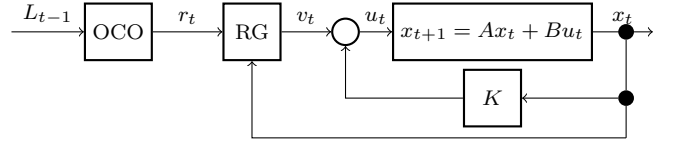

\subsection{OCO-RG scheme} \FloatBarrier

In this section, we introduce the proposed combination of OCO and a reference governor. A block diagram of the proposed approach is shown in Figure~\ref{fig:block_diagram} and our OCO-RG scheme is given in Algorithm~1. At each time $t\in\mathbb N_{\geq1}$, given access to the previous cost function $L_{t-1}$, we measure the system state $x_t$ and
\begin{enumerate}
	\item apply one projected gradient descent step with a suitable step size $\gamma>0$ in~\eqref{eq:OCO_OGD} to the optimization problem $\{\min_{r}~L^s_{t-1}(r) \quad \text{s.t. } r \in \intS\}$ in order to compute an estimate $r_t$ of the previous optimal reference $\eta_{t-1}$, compare~\eqref{eq:def_eta}. Hence, $r_t$ tracks the optimal steady state reference $\eta_{t-1}$.
	\item Next, we apply a reference governor that enforces constraint satisfaction by computing a feasible reference command $v_t$ based on the estimate $r_t$ and the measured state $x_t$ in~\eqref{eq:RG}. The constraint in~\eqref{eq:RG_opt} ensures that the error $e_t = x_t - S_Kv_t$ between $x_t$ and the steady state corresponding to $v_t$ lies in the set $\contrMAS$, which ensures that the error is contractive and constraint satisfaction by Lemma~\ref{lem:properties_contrMAS}.
	\item Finally, we apply $u_t = v_t + Kx_t$ as given in~\eqref{eq:algo_input} to the system~\eqref{eq:sys}, receive the current cost function $L_t$, and move to time step $t+1$.
\end{enumerate}

\begin{figure}
	\vspace{0pt}
	{
		\fbox{\parbox{.94\linewidth}{
				{\large \underline{Algorithm 1: OCO-RG scheme}}
				
				\vspace{7pt}
				
				\noindent Given a step size $0<\gamma\leq \frac{2}{\alpha_v+l_v}$, a stabilizing feedback $K\in\mathbb R^{m\times n}$, and an initial reference $r_0$: \\
				
				\noindent At $t=0$: Set $\alpha_0 = 1$, $v_0=r_0$ and apply $u_0 = v_0 + Kx_0$.
				
				\noindent At each time $t\in\mathbb N_{\geq 1}$:
				
				\noindent\textbf{OCO:}
				\vspace{-2ex}
				\begin{equation}
					r_t = \Pi_{\intS} \left( r_{t-1} - \gamma\nabla L^s_{t-1}(r_{t-1}) \right), \label{eq:OCO_OGD}
				\end{equation}
				\textbf{RG:}
				\begin{subequations} \label{eq:RG}
					\begin{align}
						\alpha_t = \max_{\alpha \in [0,1]} \alpha \quad
						\text{s.t. } &(v_t,x_t-S_Kv_t) \in \mathcal \contrMAS \label{eq:RG_opt} \\
						&v_t = v_{t-1} + \alpha (r_t - v_{t-1}), \label{eq:RG_update}
					\end{align}
				\end{subequations}
				\textbf{Control Input:}
				\vspace{-2ex}
				\begin{equation}
					u_t = v_t + Kx_t. \label{eq:algo_input}
				\end{equation}
		}}
	}
	\vspace{-8pt}
\end{figure}

Since we do not have access to any cost function at time $t=0$, we apply an initial reference $v_0 = r_0$ and set $\alpha_0 = 1$. In order to ensure safe operation, i.e., satisfaction of the state and input constraints at all times, we assume that the initial reference $r_0$ is feasible.
\begin{assum} \label{assump:feasible_init}
	The initial reference $r_0\in\setS$ satisfies $(r_0,x_0-S_Kr_0)\in\contrMAS$.
\end{assum}

\begin{figure}
	\begin{center}
		\begin{tikzpicture}[scale=.9]
	\clip (-1,-0.75) rectangle (7,3);
	
	
	\coordinate (center) at (0,0);
	\node[label = {[label distance = -2ex]225:{$\eta_{t-1}$}},fill=white,inner sep = 0] (eta) at (0.75,0.75) {$\times$};	
	\node[fill=white,inner sep = 0, label = {[label distance = -1ex]5:{$r_{t-1}$}}] (rt-1) at (2.5,2.55) {$\times$};
	\node[label = {[label distance = -1ex]225:{$r_{t}$}},fill=white,inner sep = 0] (rt) at (1.5,1.875) {$\times$};
	\node[inner sep = 0pt, label = {[label distance = -1ex]-25:{$v_{t-1}$}},fill=white,inner sep = 0] (vt-1) at (5,1) {$\times$};
	\node[label = {[label distance = .5ex]270:{$v_{t}$}},fill=white,inner sep = 0] (vt) at (3,1.5) {$\times$};
	
	\draw[dashed, thick] (6, 0.5) -- (4,0) -- (2.5,1) -- (4,2.5) -- (5.5,2.5) -- cycle;
	
	\node[fill=white,inner sep = 0] at (2,0) {$\intS$};	
			
	\draw[dotted,rotate around={115:(center)}] (center) ellipse (2 and 1);
	\draw[dotted,rotate around={115:(center)}] (center) ellipse (4 and 2);
	\draw[dotted,rotate around={115:(center)}] (center) ellipse (6 and 3);
	\draw[dotted,rotate around={115:(center)}] (center) ellipse (8 and 4);
	
	\draw[very thick] (0,8) -- (-1.5,4.5) -- (1.35,-0.25) -- (8,-0.75);
	\draw[thick,->] (rt-1.center) -- node[midway,above,xshift = -6ex,fill=white,inner sep = 0,yshift=1ex] {$-\gamma L^s_{t-1}(r_{t-1})$} (rt.center);
	\draw[very thin] (vt.center) -- (rt.center);
	\draw[thick, ->] (vt-1.center) -- node[midway, above, xshift = 3ex, yshift = 1ex,fill=white,inner sep = 0] {$\alpha_t(r_t - v_{t-1})$} (vt.center);	
\end{tikzpicture}
	\end{center}
	\vspace{-5ex}
	\caption{Schematic illustration of the OCO-RG scheme. The previous cost function $L^s_{t-1}$ is represented by its level sets (dotted). The thick line represents the boundary of the set $\intS$. The dashed line represents the constraint $(v_t,x_t-S_Kv_t)\in\contrMAS$ in~\eqref{eq:RG_opt}.
	}
	\label{fig:schematic}
	\vspace{0ex}
\end{figure}
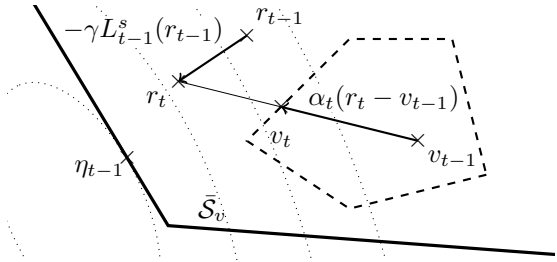

The proposed algorithm is illustrated in Figure~\ref{fig:schematic}. Note that $r_t \in \intS$ for all $t\in\mathbb N_{\geq0}$ due to the projection in~\eqref{eq:OCO_OGD} and Assumption~\ref{assump:feasible_init}. Hence, $(r_t,S_Kr_t)$ is a feasible steady state of system~\eqref{eq:sys2} at all times. 

At each time~$t$, we have to compute one gradient step and a projection in~\eqref{eq:OCO_OGD}, and solve one scalar optimization problem in~\eqref{eq:RG}. The projection can be done efficiently if the set of all feasible steady states~$\intS$, which is typically low-dimensional, has a simple structure. Thus, Algorithm~1 has a particularly low computational complexity if the projection onto the set $\intS$ can be solved efficiently.

\section{Theoretical analysis} \label{sec:results}

First, we show that the proposed OCO-RG scheme is recursively feasible, i.e., the optimization problem in~\eqref{eq:RG} has a solution at all times $t\in\mathbb N_{\geq1}$ and the algorithm is well-defined. Moreover, we show that the constraints~\eqref{eq:constraints} are satisfied at all times $t\in\mathbb N_{\geq0}$.

\begin{lem} \label{lem:rec_feas}
	Let Assumptions~\ref{assump:sys},~\ref{assump:Y}, and~\ref{assump:feasible_init} hold. Algorithm 1 is recursively feasible and $y_t \in \setY$ for all $t\in\mathbb N_{\geq0}$.
\end{lem}
\begin{pf}
	Note that $v_t\in\intS\subseteq\setS$ for all $t\geq\mathbb N_{\geq0}$, because $v_0 \in \intS$ and $r_t\in\intS$ for all $t\in\mathbb N_{\geq0}$ by~\eqref{eq:OCO_OGD} and Assumption~\ref{assump:feasible_init}. Then, Lemma~\ref{lem:properties_contrMAS}(i) implies that, if there exists a feasible reference input $v_t$ satisfying the constraint~\eqref{eq:RG_opt} at time~$t$, then $v_{t+1} = v_t$ and, hence, $\alpha_{t+1}=0$ is a feasible solution of~\eqref{eq:RG} at time $t+1$. Thus, Algorithm~1 is recursively feasible. Moreover, constraint satisfaction is guaranteed by Lemma~\ref{lem:properties_contrMAS}(ii) for all $t\in\mathbb N_{\geq0}$ because Algorithm~1 is recursively feasible. \hfill $\square$
\end{pf}

Having established that the proposed OCO-RG scheme is well-defined at all times, we analyze the closed-loop performance of Algorithm~1. In order to establish a bound on the dynamic regret, we first prove that the modified reference $v_t$ is moved towards $r_t$ at all times.

\begin{lem} \label{lem:alpha_eps}
	Suppose Assumptions~\ref{assump:sys},~\ref{assump:Y}, and~\ref{assump:feasible_init} are satisfied. There exists $\epsilon\in(0,1]$ such that $\alpha_t \geq \epsilon$ for all $t\in\mathbb N_{\geq0}$.
\end{lem}
\begin{pf}
	Fix any $t\in\mathbb N_{\geq1}$. Assumption~\ref{assump:feasible_init} and Lemma~\ref{lem:rec_feas} imply $(v_{t-1},x_{t-1}-S_Kv_{t-1})\in\contrMAS$ and, thus,
	\begin{align}
		&(v_{t-1}, x_{t-1} - S_Kv_{t-1}) \in \contrMAS \nonumber \\
		\Leftrightarrow~&e_{t-1} \in \setE(v_{t-1}) \refarrow{\eqref{eq:lambda_contr_error}} A_Ke_{t-1} \in \lambda \setE(v_{t-1}) \nonumber \\
		\Leftrightarrow~&\left(v_{t-1},\frac{1}{\lambda}A_K(x_{t-1} - S_Kv_{t-1}) \right) \in 	\contrMAS \nonumber \\
		\Leftrightarrow~&\left(v_{t-1},\frac{1}{\lambda} (x_t - S_Kv_{t-1}) \right) \in \contrMAS, \label{eq:aux_lemma}
	\end{align}
	because $A_KS_Kv_{t-1} = S_Kv_{t-1}-Bv_{t-1}$. Let $\intY$ be a compact subset of $\int\setY$ such that $(CS_K+D)\intS \subseteq \intY$, and let $\intMAS$ be the MAS of~\eqref{eq:sys_MAS} with $\setY$ replaced by $\intY$ in~\eqref{eq:sys_MAS_constr}. Then, $\intMAS\subseteq\int\contrMAS$ by definition, and $(r,0)\in\intMAS$ for all $r \in \intS$ because $r \in \intS \Leftrightarrow C\cdot0+(CS_K+D)r \in \intY \Leftrightarrow (r,0) \in \intMAS$. Hence, there exists $\delta>0$ such that $\left(r + r^\delta,-S_Kr^\delta \right) \in \contrMAS$ for all $r\in\intS$ and $r^\delta \in \delta \mathcal B$, where $\mathcal B\subseteq \mathbb R^{m}$ is the unit ball. We proceed by a case distinction.
	
	\textit{Case 1:} $\norm{r_t - v_{t-1}} \geq (1-\lambda)\delta$. Let $r_t^\delta := \delta \frac{r_t - v_{t-1}}{\norm{r_t - v_{t-1}}} \in \delta \mathcal B$ and $\alpha^c_t := \frac{(1-\lambda)\delta}{\norm{r_t-v_{t-1}}}\in(0,1]$. Since $v_{t-1} \in \intS$, we have $(v_{t-1}+r^\delta_t,-S_Kr^\delta_t) \in \contrMAS$ as shown above. Applying a convex combination with~\eqref{eq:aux_lemma} yields
	\begin{align*}
		&\lambda \left( v_{t-1},\frac{1}{\lambda}(x_{t} - S_Kv_{t-1}) \right) \\
		&\quad + (1-\lambda)
		\left( v_{t-1} + r_t^\delta, -S_K r_t^\delta \right) \in \contrMAS \\
		\Leftrightarrow &\left( v_{t-1} + \alpha^c_t(r_t-v_{t-1}), \right. \\
		&\quad \left. x_{t} - S_K \left(v_{t-1} + \alpha^c_t (r_t-v_{t-1}) \right) \right) \in \contrMAS.
	\end{align*}
	Comparing the above inclusion to the constraints in~\eqref{eq:RG}, we get that $v_t^c = v_{t-1} + \alpha_t^c (r_t - v_{t-1})$ is a feasible candidate solution at time $t$ because \mbox{$0<\alpha_t^c\leq1$.} Since $\alpha_t$ is maximized in~\eqref{eq:RG}, we get
	\[
		1\geq\alpha_t \geq \alpha_t^c = \frac{(1-\lambda)\delta}{\norm{r_t - v_{t-1}}} \geq (1-\lambda)\frac{\delta}{\Delta} =: \epsilon > 0,
	\]
	where $\Delta>0$ is a finite constant satisfying $\Delta \geq \max_{\nu_1,\nu_2\in\intS} \norm{\nu_1-\nu_2}$, which exists by compactness of $\intS$.
	
	\textit{Case 2:} $\norm{r_t - v_{t-1}} < (1-\lambda)\delta$. Let $r^\delta_t := \frac{r_t - v_{t-1}}{1-\lambda} \in \delta \mathcal B$, which again implies $(v_{t-1}+r^\delta_t,-S_Kr^\delta_t) \in \contrMAS$. Applying the same convex combination as before, we get $\left( r_t, x_{t} - S_K r_t \right) \in \contrMAS$,	i.e., $v_t = r_t$ and $\alpha_t = 1 \geq \epsilon$ is a feasible solution to~\eqref{eq:RG_opt} at time~$t$.
	
	Combining the results above, we have $\alpha_t \geq \epsilon$ for all $t\in\mathbb N_{\geq1}$. Therefore, $\alpha_0 = 1$ concludes the proof. \hfill $\square$
\end{pf}

Finally, we are ready to analyze closed-loop performance.

\begin{thm} \label{thm:regret_bound}
	Let Assumptions~\ref{assump:sys}-\ref{assump:L}, and~\ref{assump:feasible_init} be satisfied. The closed loop achieves
	\begin{equation}
		\mathcal R \leq K_0  + K_\eta \sum_{t=1}^T \norm{\eta_t - \eta_{t-1}}, \label{eq:regret_bound}
	\end{equation}
	where $K_0 = K_{0,\theta} \norm{x_0 - \theta_0} + K_{0,\eta} \norm{v_0 - \eta_0}$ and \linebreak
	$K_{0,\theta},K_{0,\eta}, K_\eta > 0$ are constants independent of $T$.
\end{thm}
\begin{pf}
	Lipschitz continuity from Assumption~\ref{assump:L} together with $\theta_t = S_K\eta_t$ yields
	\begin{align*}
		\mathcal R &\refeq{\eqref{eq:def_regret}} \sum_{t=0}^T L_t(v_t + Kx_t,x_t) - L_t(\eta_t+K\theta_t,\theta_t) \\
		&\leq l_L\sum_{t=0}^T \norm{\begin{bmatrix} v_t + Kx_t \\ x_t \end{bmatrix} - \begin{bmatrix} \eta_t + K\theta_t \\ \theta_t \end{bmatrix}} \\
		&= l_L \sum_{t=0}^T \norm{\begin{bmatrix} K(x_t - S_Kv_t)\\ x_t - S_Kv_t \end{bmatrix} + \begin{bmatrix} (KS_K+I_m)(v_t - \eta_t)\\ S_K(v_t - \eta_t) \end{bmatrix}} \\
		&\leq k_{x}\sum_{t=0}^T \norm{x_t - S_Kv_t} + k_v \sum_{t=0}^T \norm{v_t - \eta_{t}}, \numberthis \label{eq:regret_bound_prelim}
	\end{align*}
	where we defined $k_v := l_L\norm{[(KS_K+I_m)^\top, ~S_K^\top]^\top}$ and $k_x := l_L\norm{[K^\top, ~I_n]^\top}$.
	In the following, we proceed to bound the sums in \eqref{eq:regret_bound_prelim} separately. First, note that the optimal steady-state input $\eta_t$, its estimate $r_t$, and the modified reference input $v_t$ are only defined for $t \in \mathbb N_{\geq0}$. Thus, we set without loss of generality $\eta_{-1} = v_{-1} = r_{-1} = r_0 = v_0$. Moreover, we make use of the following standard result from convex optimization.
	
	\begin{lem} \cite[Theorem 2.2.14]{Nesterov18}
		Suppose Assumption~\ref{assump:Y} and~\ref{assump:L} are satisfied and let $\gamma \in (0, \frac{2}{\alpha_v+l_v}]$. Then,
		\begin{equation}
			\norm{\Pi_{\intS} \left( r - \gamma \nabla L^s_t(r) \right)-\eta_t} \leq \kappa \norm{r - \eta_t} \label{eq:gradient_contraction}
		\end{equation}
		holds for any $r\in\intS$, where $\kappa = 1-\gamma\alpha_v \in [0,1)$.
	\end{lem}
	Using \eqref{eq:gradient_contraction}, the triangle inequality, and $r_{-1}=\eta_{-1}$, we get
	\begin{equation*}
		\sum_{t=0}^T \norm{r_{t} - \eta_{t-1}} \refleq{\eqref{eq:gradient_contraction}} \kappa \sum_{t=0}^{T} \norm{r_t - \eta_{t-1}} + \kappa \sum_{t=0}^{T} \norm{\eta_t - \eta_{t-1}}.
	\end{equation*}
	Rearranging yields
	\begin{equation}
		\sum_{t=0}^T \norm{r_{t} - \eta_{t-1}} \leq \frac{\kappa}{1-\kappa} \sum_{t=0}^{T-1} \norm{\eta_t - \eta_{t-1}}. \label{eq:regret_OGD}
	\end{equation}
	Next, using the triangle inequality we get
	\begin{align*}
		&\sum_{t=0}^T \norm{v_t - \eta_t} \refleq{\eqref{eq:RG_update}} \sum_{t=0}^T (1-\alpha_t)\norm{v_{t-1} - \eta_{t-1}} \\
		&\qquad  + \sum_{t=0}^T \alpha_t \norm{r_t - \eta_{t-1}} + \sum_{t=0}^T\norm{\eta_t - \eta_{t-1}} \\
		&\refleq{\eqref{eq:regret_OGD}} (1-\epsilon) \sum_{t=0}^T \norm{v_t - \eta_t} + \frac{1}{1-\kappa}\sum_{t=0}^T \norm{\eta_t - \eta_{t-1}},
	\end{align*} 
	where we used $1\geq \alpha_t \geq \epsilon>0$ and $v_{-1} = \eta_{-1}$. Again, rearranging yields
	\begin{equation}
		\sum_{t=0}^T \norm{v_t - \eta_t} \leq \frac{1}{\epsilon_\kappa} \sum_{t=0}^T \norm{\eta_t - \eta_{t-1}}, \label{eq:regret_bound_1}
	\end{equation}
	where $\epsilon_\kappa := \epsilon(1-\kappa)$. Moreover, $v_{-1} = \eta_{-1}$ and~\eqref{eq:regret_bound_1} imply
	\begin{align*}
		\sum_{t=0}^T \norm{v_t - v_{t-1}} &\leq 2\sum_{t=0}^T \norm{v_t - \eta_t} + \sum_{t=0}^T \norm{\eta_t - \eta_{t-1}} \\
		&\refleq{\eqref{eq:regret_bound_1}} \frac{2+\epsilon_\kappa}{\epsilon_\kappa} \sum_{t=0}^T \norm{\eta_t - \eta_{t-1}}. \numberthis \label{eq:vt-vt-1}
	\end{align*}
	It remains to bound the first sum in~\eqref{eq:regret_bound_prelim}. First, consider the error dynamics $e_t = x_t - S_Kv_t$ given by $e_t = A_K e_{t-1} + S_K(v_{t-1} - v_t)$. Applying this equation repeatedly yields
	\begin{equation}
		e_t = A_K^t e_0 + \sum_{i=0}^{t-1} A_K^i S_K (v_{t-i-1} - v_{t-i}). \label{eq:error_dynamics}
	\end{equation}
	Since $A_K$ is Schur stable, there exist constants $c\geq 1$, $\sigma\in(0,1)$ such that $\norm{A_K^t}\leq c\sigma^t$. Hence, applying the closed-loop error dynamics dynamics~\eqref{eq:error_dynamics} yields
	\begin{align*}
		&\sum_{t=0}^T \norm{x_t - S_Kv_t} \refleq{\eqref{eq:error_dynamics}} \sum_{t=0}^T \norm{A_K^t} \norm{e_0} \\
		&\qquad + \norm{S_K} \sum_{t=0}^T \sum_{i=0}^{t-1} \norm{A_K^i} \norm{v_{t-i-1} - v_{t-i}} \\
		\leq &\frac{c}{1-\sigma} \norm{e_0} + \norm{S_K} \sum_{t=0}^T \left( \norm{v_t - v_{t-1}} \sum_{i=0}^{T-1} \norm{A_K^i} \right) \\
		\refleq{\eqref{eq:vt-vt-1}} &\frac{c}{1-\sigma} \norm{x_0 - \theta_0} + \frac{c}{1-\sigma}\norm{S_K} \norm{v_0 - \eta_0} \\
		&\qquad + \frac{c}{1-\sigma} \norm{S_K} \frac{2+\epsilon_\kappa}{\epsilon_\kappa} \sum_{t=0}^T \norm{\eta_t - \eta_{t-1}}. \numberthis \label{eq:regret_bound_2}
	\end{align*}
	Inserting \eqref{eq:regret_bound_1} and \eqref{eq:regret_bound_2} into \eqref{eq:regret_bound_prelim}, and using $\eta_{-1}=v_0$ proves the result~\eqref{eq:regret_bound}. \hfill $\square$
\end{pf}

\section{Numerical example} \label{sec:example}

In this section, we illustrate the performance of the proposed OCO-RG scheme by numerical simulation. We consider the problem of tracking a time-varying and a priori unknown reference while minimizing control effort. In particular, the linear system~\eqref{eq:sys} is generated randomly by sampling each entry of $A \in \mathbb{R}^{5\times5}$ from a uniform distribution over the interval $[-1,1]$. We obtained an unstable system with $\rho(A)\approx1.62$ and set $B=\begin{bmatrix} 0 & \dots & 0 & 1 \end{bmatrix}^\top$. We consider box constraints of the form $|x_i|\leq1$, $i\in\{1,\dots,5\}$, and $|u|\leq1$. The stabilizing controller $K$ is chosen such that the eigenvalues of $A_K$ are given by $\text{eig}(A)=\{0.1,0.15,\dots,0.3\}$. Finally, we let $\lambda = 0.95$, $\intS = 0.95\setS \subseteq \int \setS$, and compute $\contrMAS$ using the multi-parametric toolbox~\citep{MPT3}. The cost functions are given by $L_t(u,x) = \frac{1}{2}\norm{x - \bar x_t}^2 + \frac{q_t}{2} \norm{u_t}^2$, where $\bar x_t = \bar z_t + 0.2\sin(\frac{\pi}{100}t)$, and both $\bar z_t \in [-1,1]$ and $q_t \in [0,2]$ are a priori unknown. More specifically, at each time $t$, we change $q_t$ with a probability of $1\%$ by uniformly sampling it from the interval $[0,2]$ and keep it constant otherwise. We apply the same procedure to $\bar z_t \in [-1,1]$. Finally, we set $\gamma = 0.1$ and initialize the system and the algorithm with $x_0=0_n$, and $r_0=0$. The results are illustrated in Figure~\ref{fig:sim}. The top plot of Figure~\ref{fig:sim} shows the first state of the closed-loop system $x_1$ together with the optimal steady state $\theta_1$ and the corresponding constraints. It can be seen that the closed loop follows the optimal steady state closely, for sudden changes as well as for slow, continuous changes induced by the sine term in the definition of $\bar x_t$. Moreover, the middle plot of Figure~\ref{fig:sim} illustrates the optimal steady-state input $\eta_t+K\theta_t$, the estimate $r_t+Kx_t$, the applied input $u_t=v_t+Kx_t$, and the input constraints. Since the reference command $r_t$ together with the stabilizing feedback $Kx_t$ would violate the input constraints, the reference governor modifies the reference such that $u_t$ satisfies the constraints at all times. The bottom plot of Figure~\ref{fig:sim} shows the parameter $\alpha_t$. As proven in Lemma~\ref{lem:alpha_eps}, $\alpha_t>0$ at all times $t$. More specifically, in this simulation the lowest value of $\alpha_t$ is approximately $6\cdot10^{-3}$. The maximum computation time of the proposed OCO-RG scheme\footnote{The simulations were performed on a standard laptop (Intel Core i9 with 2.6 GHz and 16 GB RAM under Windows 10) in Matlab.} averaged over $100$ trials and different realizations of the cost function is approximately $66\mu$s per time step.

\setlength\breite{.4\textwidth}
\setlength\hohe{22ex}
\begin{figure}
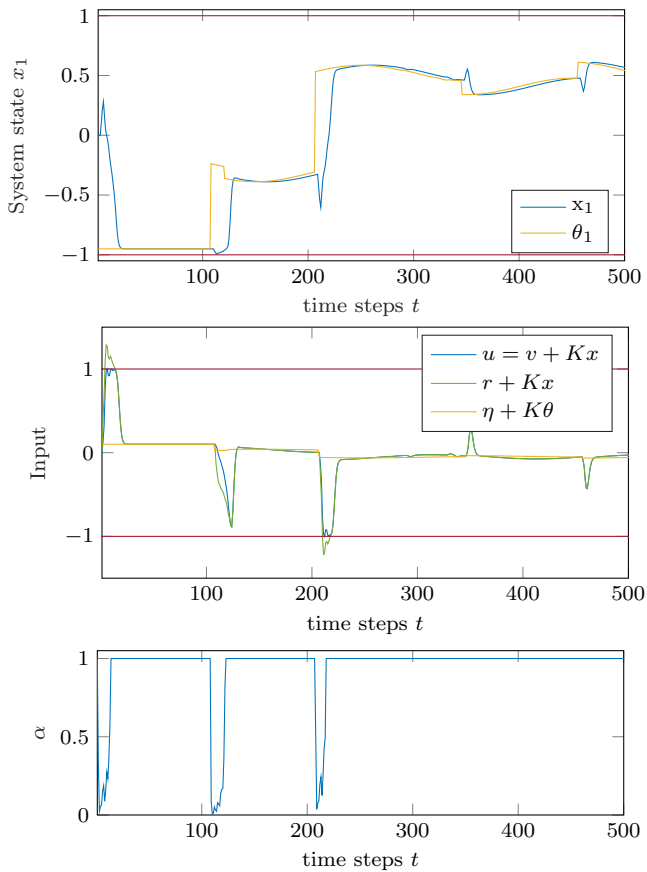
 
	\begin{center}
		\small
		\input{figures/simulation_221110/sim_state.tex}

		\small
		\input{figures/simulation_221110/sim_input.tex}
		
		\setlength\hohe{18ex}
		\small
		\input{figures/simulation_221110/sim_alpha.tex}
	\end{center}
	\vspace{-2ex}
	\caption{Closed-loop trajectories of the simulation example.}
	\label{fig:sim}
\end{figure}

\section{Conclusion} \label{sec:conclusion}

In this paper, we propose an algorithm for controlling dynamical systems subject to time-varying and a priori unknown cost functions as well as pointwise in time state and input constraints by combining the online convex optimization framework with a reference governor. In particular, we make use of a $\lambda$-contractive set to ensure constraint satisfaction at all times as well as a sufficient convergence rate for proving that the closed loop's dynamic regret is bounded linearly in the variation of the cost functions.
Future works includes obtaining theoretical guarantees for the practically important case of disturbances. 

\footnotesize
\bibliography{bib}
\end{document}